\documentstyle[12pt]{report}

\newcommand{\nonumsection}[1] {\vspace{12pt}\noindent{\bf #1}
        \par\vspace{5pt}}

\def\thebibliography#1{\nonumsection{\large \it References}\list
  {[\arabic{enumi}]}{\settowidth\labelwidth{[#1]}\leftmargin\labelwidth
    \advance\leftmargin\labelsep
    \usecounter{enumi}}
    \def\newblock{\hskip .11em plus .33em minus .07em}
    \sloppy\clubpenalty4000\widowpenalty4000}

\newcommand{\be}{\begin{eqnarray}}
\newcommand{\ee}{\end{eqnarray}}

\begin{document}
\baselineskip24pt

\rightline{UNITU-THEP-14/1995}
\rightline{September 1995}
\rightline{hep-ph/9509396}

\vskip 1.5truecm
\centerline{\Large\bf Radiative Decays of Hyperons in the}
\vskip 0.3cm
\centerline{\Large\bf Skyrme Model: E2/M1 Transitions
Ratios$^\dagger$}
\vskip 1.0cm
\centerline{Abdellatif Abada$^{\S}$, Herbert Weigel$^\ddagger$,
and Hugo Reinhardt}
\vskip 0.3cm
\centerline{Institute for Theoretical Physics}
\centerline{T\"ubingen University}
\centerline{Auf der Morgenstelle 14}
\centerline{D-72076 T\"ubingen, Germany}
\vskip 2.0cm
\centerline{\bf ABSTRACT}
\vskip 0.5cm
\noindent
We study the radiative decays of $J^\pi=\frac{3}{2}^+$ baryons in
the framework of the SU(3) collective approach to the Skyrme model.
We present the predictions for the decay widths and the
corresponding $E2/M1$ ratios. We find that all considered ratios are
negative and of the order of a few percent only. We discuss the
effects of flavor symmetry breaking and compare our
results to those obtained in related models.

\vfill

\noindent
------------\hfil\break
\footnotesize
\noindent
$^\dagger $ Supported by the Deutsche Forschungsgemeinschaft
(DFG) under contract Re 856/2-2.
\newline
\noindent
$^{\S}$ Address after Oct. $1^{\rm st}$, 1995: Department of
Physics and Astronomy, University of Manchester, Manchester
M13 9PL, UK.
\newline
\noindent
$^\ddagger$ Supported by a Habilitanden--scholarship of the DFG.
\newpage

\normalsize
\baselineskip 22pt

\stepcounter{chapter}
\leftline{\large\it 1. Introduction}
\smallskip

At present, only few data are available concerning the electromagnetic
decays of hyperons. The most prominent one certainly is the reaction
$\Delta\to N\gamma$.  Recently, the ratio of the electric quadrupole
($E2$) to the magnetic dipole ($M1$) amplitude has been reanalyzed to
be $E2/M1= (-2.5\pm0.2)\%$ from a $\pi^{0(+)}$--photoproduction
experiment performed at MAMI \cite{MAMI}\footnote{This preliminary
result contains both, resonant and non--resonant contributions.
These may add incoherently leading to an even larger (in magnitude)
ratio for the resonant piece.}. For the $J=\frac{3}{2}$ to
$J=\frac{1}{2}$ transitions, which involve strange
baryons, the empirical values for the $E2/M1$ ratios are still
unknown. Upcoming experiments at CEBAF \cite{CEBAF} and Fermilab
\cite{Fermilab} are expected to soon provide some data on these
radiative decays. However, these transitions have already been studied
within several models, which include the non--relativistic quark model
\cite{DHR,Le93}, the MIT bag model \cite{KMS}, heavy baryon chiral
perturbation theory \cite{BSS} as well as a quenched lattice
calculation \cite{Le93}. More recently, Schat {\it et al.}
\cite{SGS} presented a detailed analysis of the hyperon radiative
decays within the bound state approach \cite{CK} to the Skyrme model
\cite{SK}. In that treatment hyperons are considered as kaons bound in
the background of the static soliton field. In the Skyrme model hyperons
may alternatively be described within the SU(3) collective treatment.
In the latter approach strange degrees of freedom are incorporated as
SU(3) collective excitations of the non--strange soliton. Canonical
quantization of the collective coordinates yields a Hamiltonian,
which may be diagonalized exactly \cite{YA,Pa89} although it contains
flavor symmetry breaking pieces. This procedure provides the baryon
energies and wave--functions in the space of the collective
coordinates. In the present work we employ this collective approach
to the SU(3) Skyrme model for investigating the transitions
$B(J=\frac{3}{2})\to \gamma B\prime(J=\frac{1}{2})$. Further, we
will compare these results with those obtained in the studies
mentioned above.

\bigskip

\stepcounter{chapter}
\leftline{\large\it 2. The collective approach to the SU(3)
Skyrme model}
\smallskip

Our starting point is the non--linear realization, $U={\rm exp}(i\Phi)$,
of the pseudoscalar nonet, $\Phi$. Chirally invariant objects are
conveniently constructed by introducing the derivative of $U$ via
$\alpha_\mu=U^{\dag}\partial_\mu U$. The Skyrme model contains
the non--linear $\sigma$ model and the forth--order stabilizing term
\be
{\cal L}_{\rm S}={\rm Tr}
\left(-\frac{\tilde f_{\pi}^2}{4}\alpha_{\mu}\alpha^{\mu}+
\frac{1}{32e^2} [\alpha_{\mu},\alpha_{\nu}]
[\alpha^{\mu},\alpha^{\nu}]\right)
\label{lsk}
\ee
which are flavor symmetric. In flavor SU(3) a minimal set of symmetry
breaking terms is included \cite{We90}
\be
{\cal L}_{\rm SB}=
{\rm Tr}(T+xS)\left[
\beta^\prime(U\alpha_{\mu}\alpha^{\mu}+
\alpha_{\mu}\alpha^{\mu}U^{\dag})
+ \delta^\prime(U+U^{\dag}-2)\right].
\label{lsb}
\ee
Here $T={\rm diag}(1,1,0)$ and $S={\rm diag}(0,0,1)$ are the
projectors onto the non--strange and strange degrees of freedom,
respectively. The parameters are determined from the masses and
decay constants of the pion and the kaon. Note, that the physical
pion decay constant ($f_\pi=93{\rm MeV}$) is given by
$f_\pi^2=\tilde f_\pi^2-8\beta^\prime$. To be explicit
$\beta^\prime=-26.4{\rm MeV}^2$, $\delta^\prime=4.15\times10^{-5}
{\rm GeV}^4$ and $x-1\approx36$ measures the flavor symmetry breaking
\cite{We90}. To take proper account of the axial anomaly, the
Wess--Zumino term
\be
\Gamma_{\rm WZ}=-\frac{iN_c}{240 \pi^2} \int d^5x
\epsilon^{\mu \nu\rho\sigma\kappa}{\rm Tr}
\left(\alpha_\mu\alpha_\nu\alpha_\rho
\alpha_\sigma\alpha_\kappa\right)
\label{WZterm}
\ee
is added. For the study of electromagnetic properties
of baryons at finite momentum transfer a direct derivative
coupling to the photon field, $A_\mu$, has proven relevant
\be
{\cal L}_9=iL_9\left(\partial_\mu A_\nu -\partial_\nu A_\mu\right)
{\rm Tr}\left(\xi^{\dag}\left[\lambda_3+
\frac{1}{\sqrt3}\lambda_8\right]\xi
\left[\xi^{\dag}\alpha^\mu\alpha^\nu\xi+
\xi\alpha^\mu\alpha^\nu \xi^{\dag}\right]
\right)\ ,
\label{lagl9}
\ee
where the square root of the chiral field has been introduced,
{\it i.e.} $U=\xi^2$. In forth order chiral perturbation this term
is needed to correctly reproduce the electromagnetic pion radius
determining the dimensionless coefficient
$L_9=(6.9\pm0.7)\times10^{-3}$ \cite{GL}.
The total action is the sum
\be
\Gamma=\int d^4x\left\{{\cal L}_{\rm S}+
{\cal L}_{\rm SB}+{\cal L}_9\right\}
+\Gamma_{\rm WZ}\ .
\label{action}
\ee
The associated electromagnetic current, $J^{\rm e.m}_\mu$ is
obtained in two steps. First, the photon field is
incorporated such that the action is invariant under the local
$U_{\rm e.m.}(1)$ gauge transformation (${\cal L}_9$ already has this
property). Secondly, $J_{\rm e.m}^\mu$ is identified as the object
which couples to the photon field linearly. The resulting covariant
expression may {\it e.g.} be found in ref \cite{Pa91}.

The SU(3) collective rotational approach for the description of
the hyperons as chiral solitons employs the time dependent
meson configuration \cite{We90}
\begin{equation}
\xi(\mbox{\boldmath $r$},t)=
A(t)\xi_k(\mbox{\boldmath $r$})\xi_{\rm H}(\mbox{\boldmath $r$})
\xi_k(\mbox{\boldmath $r$}) A^{\dag}(t)
\label{coll}
\end{equation}
Here $\xi_{\rm H}({\bf r})$ refers to the hedgehog {\it ansatz}
$$
\xi_{\rm H}\left(\mbox{\boldmath $r$}\right)=
{\rm exp}\left(i\hat{\mbox{\boldmath $r$}}\cdot
\mbox{\boldmath $\tau$}F(r)/2\right),
$$
while the SU(3) matrix $A$ contains the collective coordinates.
The time dependence of $A$ is most conveniently parametrized
in terms of the eight angular velocities
$\Omega_a=-i {\rm Tr} \lambda_a A^{\dag}{\dot A}$. It is
convenient to also introduce the adjoint representation of the
collective rotation,
$D_{ab}=(1/2){\rm Tr}(\lambda_a A\lambda_b A^{\dag})$.
The kaon fields, which are induced by the collective rotation
are contained in $\xi_k={\rm exp}(iZ)$
\be
Z= W(r)d_{i\alpha\beta}\hat{\mbox{\boldmath $r$}}_i
\Omega_{\alpha}\lambda_{\beta} \ .
\label{kaonind}
\ee
As usual, the convention $i=1,2,3$ and $\alpha,\beta=4,...,7$ is
adopted. In this {\it ansatz} $\lambda_a$ and $d_{abc}$ denote
the Gell--Mann matrices and symmetric structure functions of SU(3),
respectively. Although the inclusion of these induced fields is
mandatory to satisfy the PCAC--type relation for the kaon fields
they introduce double counting effects because the overlap with the
rotation of the pion fields into the strange flavor direction,
$Z_0\sim i[\lambda_\alpha,\xi_H(\mbox{\boldmath $r$})]$,
\be
\langle Z_0 | Z \rangle \propto
\int_0^\infty r^2 dr W(r)\ {\rm sin}\frac{F(r)}{2}
\label{overlap}
\ee
vanishes for infinitely large symmetry breaking only \cite{We90}.
Later we will alternatively consider the model (\ref{action})
augmented by a Lagrange multiplier enforcing (\ref{overlap}) to
vanish.

The configuration (\ref{coll}) is then substituted into
the action yielding the Lagrangian of the collective
coordinates $L(A,\Omega_a)$. Canonical quantization of the collective
coordinates provides a linear relation between the angular
velocities and the right generators of SU(3),
$R_a=-\partial L(A,\Omega_a)/\partial\Omega_a$. For $i=1,2,3$ this
relation defines the operator for the total angular momentum
$J_i=-R_i$. Diagonalization of the associated Hamiltonian,
$H(A,R_a)=-R_a\Omega_a-L$,
generates the states corresponding to physical baryons.
Although $H(A,R_a)$ contains flavor symmetry breaking terms it
can be diagonalized exactly \cite{YA,Pa89} yielding the baryon
wave--functions in the space of the collective coordinates.

\bigskip

\stepcounter{chapter}
\leftline{\large\it 3. Radiative decays of hyperons}
\smallskip

In order to extract information about the radiative decays of the
$\frac{3}{2}^+$ baryons we need the quadrupole and monopole pieces
of the electric and magnetic form factors, respectively. The former is
extracted from the orbital angular momentum $l=2$ component of the
time component of the electromagnetic current, $J_0^{\rm e.m.}$, while
the latter is obtained from the spatial components, $J_j^{\rm e.m.}$.
It is therefore suitable to define the associated Fourier transforms
\be
\hat E(q)&=&\int d^3r\ j_2(qr)
\left(\frac{z^2}{r^2}-\frac{1}{3}\right)J_0^{\rm e.m.}
\nonumber \\
\hat M(q)&=&\frac{1}{2}\int d^3r\ j_1(qr)\epsilon_{3ij}
{\hat r}_iJ_j^{\rm e.m.} \ ,
\label{Ftrans}
\ee
where the $j_l(qr)$ denote spherical Bessel functions. Substituting
the {\it ansatz} (\ref{coll}) into $J_j^{\rm e.m.}$ \cite{Pa91} yields
the electric quadrupole operator
\be
\hat E(q)&=&-\frac{8\pi}{15\alpha^2}D_{{\rm e.m.},3}R_3
\int_0^\infty dr r^2 j_2(qr)V_0(r) \ ,
\nonumber \\
V_0(r)&=&s^2\left[f_{\pi}^2+
\frac{1}{e^2}\left(F^{\prime2}+\frac{s^2}{r^2}\right)
-8\beta^\prime c \right]
-4L_9\left(s^{\prime2}+\frac{s}{r}(rs)^{\prime\prime}
-3\frac{s^2}{r^2}\right)\ ,
\label{e2op}
\ee
where contributions carrying total angular momentum zero have
been omitted.  Derivatives of radial functions with respect to the
radial coordinate are denoted by a prime. Furthermore the
abbreviations $s={\rm sin}F$ and $c={\rm cos}F$ have been
introduced and the electromagnetic direction
$D_{{\rm e.m.},i}=D_{3i}+D_{8i}/\sqrt3$ has been indicated. The
moment of inertia $\alpha^2$ for rotation in coordinate space
appears because the angular velocity $\Omega_i$ has been
substituted by the corresponding right generators,
$R_i=-\alpha^2\Omega_i$. Similarly, the magnetic monopole operator
becomes
\be
{\hat M}(q)&=&-\frac{4\pi}{3}\int_0^\infty dr r^3 j_1(qr)
\Bigg\{V_1(r)D_{{\rm e.m.},3}-\frac{1}{\beta^2}V_2(r)
d_{3\alpha\beta}D_{{\rm e.m.},\alpha}R_\beta
+V_3(r)D_{88}D_{{\rm e.m.},3}
\nonumber \\ && \hspace{4cm}
+V_4(r)d_{3\alpha\beta}D_{{\rm e.m.},\alpha}D_{8\beta}
+\frac{\sqrt3}{2\alpha^2}B(r)D_{{\rm e.m.},8}R_3\Bigg\} \ .
\hspace{1cm}
\label{m1op}
\ee
The moment of inertia $\beta^2$ for rotation
into the strange flavor direction stems from the replacement
$R_\alpha=-\beta^2\Omega_\alpha$. Except for the contributions
of the pion--radius term (\ref{lagl9}) to $V_1$ and $V_2$
\be
V^{(L_9)}_1(r)&=&-L_9 \frac{1}{r^2}
\left[({\rm cos}2F)^{\prime\prime}+4\frac{s^2}{r^2}\right]
\nonumber \\
V^{(L_9)}_2(r)&=&\frac{4L_9}{r^2}\left\{
\left[c_2(1+c_2)(W^\prime s+Wc_2F^\prime)
-WsF^\prime(s+s_2)\right]^\prime
-\frac{2}{r^2}Wsc_2(1+c_2) \right\}
\hspace{1cm}
\label{v12l9}
\ee
the explicit expressions for the radial functions
$V_1(r),\ldots,B(r)$ in eq (\ref{m1op}) may be traced from refs
\cite{Pa91,Pa92}. In addition to the abbreviations defined after
eq (\ref{e2op}) we have introduced $s_2={\rm sin}(F/2)$ and
$c_2={\rm cos}(F/2)$. The current $J_\mu^{\rm e.m}$ is formally
identical in the two approaches I and II, {\it i.e.} in II
we omit the explicit contribution stemming from the constraint
$\langle Z_0|Z\rangle=0$.

The decay widths ($\Gamma$) for the radiative decays of the
$\frac{3}{2}^+$ baryons to $\frac{1}{2}^+$ baryons are
obtained as the appropriate matrix elements of ${\hat E}$
and ${\hat M}$
\be
\Gamma_{E2}&=&\frac{675}{8}\alpha_{\rm hf}\ q\ \left|\langle
B(\mbox{\small $\frac{1}{2}$}^+)|{\hat E}(q)|
B^\prime(\mbox{\small $\frac{3}{2}$}^+)\rangle\right|^2\ ,
\label{Ge2} \\
\Gamma_{M1}&=&18\alpha_{\rm hf}\ q\ \left|\langle
B(\mbox{\small $\frac{1}{2}$}^+)|{\hat M}(q)|
B^\prime(\mbox{\small $\frac{3}{2}$}^+)\rangle\right|^2\ ,
\label{Gm1}
\ee
where $q$ refers to the momentum of the photon in the rest frame
of the $\frac{3}{2}^+$ baryon and $\alpha_{\rm hf}=1/137$ denotes
the electromagnetic structure constant. The matrix
elements indicated in eqs (\ref{Ge2}) and (\ref{Gm1}) are computed
in the space of the collective coordinates, employing the baryon
wave--functions, which exactly diagonalize the collective Hamiltonian,
$H(A,R_a)$. This especially implies that the baryon states are not
pure octet (decouplet) states for the $\frac{1}{2}^+$ ($\frac{3}{2}^+$)
baryons but rather contain admixtures of higher dimensional SU(3)
representations. For the details of the calculational procedure using
an ``Euler angle'' decomposition for the rotation matrix $A$ we refer
the interested reader to appendix A of ref \cite{Pa91}. These analyses
also allow us to compute the desired
$E2/M1$--ratio from \cite{Wi87}
\be
\frac{E2}{M1}= \frac{5}{4}\
\frac{\langle B(\frac{1}{2}^+)|{\hat E}(q)|
B^\prime(\frac{3}{2}^+)\rangle}
{\langle B(\frac{1}{2}^+)|{\hat M}(q)|
B^\prime(\frac{3}{2}^+)\rangle} \ .
\label{E2M1}
\ee

\bigskip

\stepcounter{chapter}
\leftline{\large\it 4. Numerical results}
\smallskip

The Skyrme parameter $e$ is fixed by optimizing the model predictions
for the baryon mass differences. For this purpose we determine the
minimum of $\chi=(1/7)[\sum (\triangle M_{\rm pred} -
\triangle M_{\rm expt})^2]^{1/2}$  as a function of $e$.
The sum goes over the seven mass differences $\triangle M=$
$M_\Lambda-M_N$, $M_\Sigma-M_N$,..., $M_\Omega-M_N$. This results in
$e=4.0$ with $\chi=13.0{\rm MeV}$, while a computation which includes
the constraint, $\langle Z_0|Z\rangle=0$, requires $e=3.9$ yielding
$\chi=12.4{\rm MeV}$. In the following we will refer to these two
approaches by I and II, respectively. For later reference we also
quote the model predictions for the magnetic moments \cite{Pa91}
of the nucleon, $\mu_p=2.01\ (1.98)$ and $\mu_n=-1.53 (1.57)$ for the
treatment I (II). The changes caused by the constraint are obviously
minor and can easily be compensated by a slight variation of the
Skyrme parameter.  In both cases the experimental values
($\mu_p=2.79,\ \mu_n=-1.91$) are underestimated. It should also be
mentioned that the pion--radius term (\ref{lagl9}) does not contribute
to the magnetic moments because the corresponding current is a total
derivative.

We now turn to the primary issue of this paper, the predictions
for the radiative decays of the $\mbox{\small $\frac{3}{2}$}^+$
baryons. Our results are summarized in table \ref{ta_width}.
\begin{table}[t]
{\footnotesize
\caption{\label{ta_width}The predictions for the total decay widths
$\Gamma_{\rm tot}=\Gamma_{M1}+\Gamma_{E2}$ (in keV) and ratios $E2/M1$
(in percent) in the collective approach to the Skyrme model. The
treatments I and II are explained in the text. The data in
parentheses refer to the ratios $E2/M1$ being rescaled by the
proton magnetic moment. Also given are the non--relativistic
quark model (QM, \protect\cite{DHR,Le93}) and quenched
lattice (Lat., \protect\cite{Le93}) predictions for the total
widths. Note that the latter data are normalized to reproduce the
magnetic moment of the proton.}
{}~
\newline
\centerline{\tenrm
\begin{tabular}{l | c c | c c | c c | c c| c |c}
& \multicolumn{4}{c|}{I}
& \multicolumn{4}{c|}{II}
& QM & Lat. \\
\cline{2-11}
& \multicolumn{2}{c|}{$L_9=0$}
& \multicolumn{2}{c|}{$L_9=6.9\times10^{-3}$}
& \multicolumn{2}{c|}{$L_9=0$}
& \multicolumn{2}{c|}{$L_9=6.9\times10^{-3}$} & & \\
 & $\Gamma_{\rm tot}$ & $E2/M1$
 & $\Gamma_{\rm tot}$ & $E2/M1$
 & $\Gamma_{\rm tot}$ & $E2/M1$
 & $\Gamma_{\rm tot}$ & $E2/M1$
 & $\Gamma_{\rm tot}$  & $\Gamma_{\rm tot}$ \\
\hline
$\Delta \to \gamma N $ &
339 & -3.1(-2.3) & 313 & -3.7(-2.7) &
348 & -3.1(-2.2) & 322 & -3.7(-2.6) &
330 & 430 \\
$\Sigma^{*0} \to \gamma \Lambda$  &
195 & -3.2(-2.4) & 180 & -3.8(-2.8) &
209 & -3.1(-2.2) & 194 & -3.7(-2.6) &
232 & --- \\
$\Sigma^{*-} \to \gamma \Sigma^{-}$  &
1  & -6.2(-4.6) &  1  & -7.3(-5.3) &
2  & -3.7(-2.6) &  2  & -4.3(-3.1) &
2 & 3 \\
$\Sigma^{*0} \to \gamma \Sigma^{0}$  &
16  & -1.3(-1.0) &  15  & -1.5(-1.1) &
12  & -1.7(-1.2) &  12  & -1.9(-1.4) &
 18 &  17 \\
$\Sigma^{*+} \to \gamma \Sigma^{+}$  &
81  & -1.9(-1.4) & 78  & -2.2(-1.6) &
74  & -2.0(-1.5) & 71  & -2.3(-1.7) &
100 & 100 \\
$\Xi^{*-} \to \gamma \Xi^{-}$  &
3  & -5.2(-3.8) & 3   & -6.1(-4.5) &
5  & -3.6(-2.6) & 4   & -4.3(-3.0) &
3 & 4 \\
$\Xi^{*0} \to \gamma \Xi^{0}$  &
120 & -2.1(-1.6) & 115 & -2.4(-1.8) &
111 & -2.2(-1.6) & 108 & -2.6(-1.8) &
137 & 129 \\
\end{tabular}
}}
\end{table}
All considered $E2/M1$--ratios are found to be negative and of the
order of a few percent only. Furthermore the different treatments
(I, II) do not cause significant changes. The inclusion of the
pion radius term (\ref{lagl9}) tends to lower the $M1$ partial
width leading to larger $E2/M1$--ratios. We have already noted that
the proton magnetic moment is predicted too low. This motivates the
scaling\footnote{This treatment may be considered as an approximation
to possible $1/N_C$ corrections \cite{Da94}. However, in contrast to
the simple scaling, these corrections can in principle vary with the
momentum transfer $q$. Fortunately, only small momentum transfers are
involved in the radiative hyperon decays considered.}
$E2/M1\rightarrow E2/M1\times(\mu_p^{\rm pred}/\mu_p^{\rm expt})$,
which approximately causes a reduction by a factor 0.7. With this
procedure the model reproduces the newest data ($-2.5\pm0.2$)
for the $E2/M1$--ratio of the process $\Delta\rightarrow \gamma N$
\cite{MAMI}. For the widths this scaling, which also has been performed
in the quenched lattice computation of ref \cite{Le93}, results in
values approximately twice as large as those displayed in table
\ref{ta_width}, {\it e.g.}
$\Gamma_{\Delta\to\gamma N}\approx650{\rm keV}$. This is similar to
the {\it particle data group} estimate of $660\ldots730{\rm keV}$ for
the width of the decay $\Delta\to\gamma N$ \cite{PDG}.

For the proceeding discussion it is convenient to categorize the
radiative decays according to the magnitude of their widths into
large (l): $\Delta\to\gamma N$, $\Sigma^{*0}\to\gamma\Lambda$,
$\Xi^{*0}\to\gamma\Xi^0$, moderate (m): $\Sigma^{*+}\to\gamma\Sigma^+$,
$\Sigma^{*0}\to\gamma\Sigma^0$ and tiny (t): $\Xi^{*-}\to\gamma\Xi^-$,
$\Sigma^{*-}\to\gamma\Sigma^-$.

We have considered the two limiting cases of flavor symmetric and
strongly distorted baryon wave functions. In the former case the $E2$
transition matrix elements for the t--type reactions vanish
identically. In this limit, together with the omission of the symmetry
breakers $V_3$ and $V_4$ in eq (\ref{m1op}), we also observe vanishing
$M1$ transition matrix elements for these reactions. Of course, this
just reflects the $U$--spin selection rule of the flavor
symmetric formulation \cite{Li73}. Although the use of SU(3)
symmetric baryon wave--functions results in widths, which are
up 30\% smaller than those presented in table \ref{ta_width},
a small deviation from flavor symmetric wave--functions yields already
results similar to those shown in table \ref{ta_width}. In the
large symmetry breaking limit we find that those $E2$ transition
matrix elements, which do not change the baryon isospin, tend to
be proportional to the corresponding isospin projection. Actually
this feature is similarly observed in the bound state computation
\cite{SGS}. However, in the collective treatment this proportionality
is only slowly approached with increasing symmetry breaking in the
baryon wave--functions, {\it i.e.} the isoscalar component decreases
only slowly with increasing symmetry breaking. Quantitatively the
effects of symmetry breaking can be investigated by varying the
symmetry breaking parameter $x$ in eq (\ref{lsb}). Even for values as
large as $x\approx100$ a 30\% deviation from these proportionalities
is obtained. In the realistic case ($x\approx37$) the isoscalar
contribution is still sizable leading to the intermediate situation
where {\it e.g.} $\langle\Xi^-|{\hat E}(q)|\Xi^{*-}\rangle
\approx -(1/3)\langle \Xi^0|{\hat E}(q)|\Xi^{*0}\rangle$. In a
wide range of the symmetry breaking parameter $x=20\ldots50$ the
total widths of the l--type reactions exhibit almost no variation
($<5$\%). Also the absolute change of the m--type widths is only
about $10{\rm keV}$. The t--type widths may increase by a factor
two in this range, however, these are small in any event. We
therefore conclude that the flavor symmetry breaking has no
significant impact on the predictions for the radiative hyperon
decays. This is in contrast to other quantities, especially
those which are related to the strangeness content of the nucleon
\cite{We95}.

The largest decay width for reactions involving strange baryons is
obtained for $\Gamma_{\Sigma^{*0}\to\gamma\Lambda}\approx240{\rm keV}$.
This is similar to other model predictions like the bound state approach
to the Skyrme model \cite{SGS} or the non--relativistic quark model
\cite{KMS}. Our results for
$\Gamma_{\Xi^{*0} \to \gamma \Xi^0}$ $\approx110{\rm keV}$ and
$\Gamma_{\Sigma^{*+} \to \gamma \Sigma^+}\approx80{\rm keV}$
are also of the same order as these model calculations. Although
the absolute values for the decay widths predicted by the bound
state method exhibit some parameter dependencies \cite{SGS}, the
pattern
\be
\Gamma_{\Sigma^{*0}\to\gamma \Lambda}>
\Gamma_{\Xi^{*0}\to\gamma\Xi^0}>
\Gamma_{\Sigma^{*+}\to\gamma\Sigma^+}\gg
\Gamma_{\Sigma^{*0}\to\gamma\Sigma^0}\gg
\Gamma_{\Xi^{*-}\to\gamma\Xi^-}\approx
\Gamma_{\Sigma^{*-}\to\gamma\Sigma^-}\approx0
\label{pattern}
\ee
is recovered in the collective treatment. On the whole our
predictions for the widths tend to be slightly smaller than those
of the non--relativistic quark model or the quenched lattice
calculation.

\bigskip

\stepcounter{chapter}
\leftline{\large\it 5. Conclusion}
\smallskip

We have computed the decay widths for the radiative decays of the
$\frac{3}{2}^+$ baryons in the framework of the collective approach
to the SU(3) Skyrme model by separately evaluating the magnetic
dipole ($M1$) and electric quadrupole ($E2$) transition matrix
elements. The total decay widths have been found to be strongly
dominated by the $M1$ contribution yielding $E2/M1$ ratios which
are of the order of a few percent only. Hence the quadrupole
deformation of the baryons is predicted to be small in the
collective approach to the SU(3) Skyrme model. All these ratios
are predicted to be negative. The resulting decay widths agree
reasonably not only with predictions of the bound state
approach to the Skyrme model \cite{SGS} but also with those
obtained within other models of the baryon \cite{DHR,KMS,SGS}.

We have observed that these transition matrix elements are not
very sensitive to flavor symmetry breaking. This naturally
explains why even in the realistic case, where the baryon
wave--functions significantly deviate from pure octet and decouplet
states, the $U$--spin selection rule \cite{Li73} is almost exactly
reproduced. As for the magnetic moments of the $\frac{1}{2}^+$ baryons
the collective approach approximately satisfies the relations,
which reflect the U--spin symmetry \cite{Sch92}. Since a treatment,
which incorporates the flavor orientation in the stationary
condition for the chiral angle, predicts the experimentally
demanded deviations from the U--spin relations \cite{Sch92}, one
may speculate whether this treatment yields significantly different
results for the transition matrix elements.

In contrast to the bound state model \cite{SGS} we observe a
non--vanishing isoscalar contribution to the $E2$ transition matrix
elements. Only in the unrealistic case of infinitely large flavor
symmetry breaking these matrix elements are proportional to the
isospin projection. Actually this is similar to the situation for
the quadrupole moments of the $\frac{3}{2}^+$ baryons. In the bound
state approach these moments are found to be proportional to the
isospin projection \cite{Oh95a}. In the collective approach this
proportionality only appears in the large symmetry breaking limit,
while for small breaking the quadrupole moments happen to be linked
to the baryon charge \cite{Kr94}.

\end{document}